\documentclass[a4paper,amsmath,amssymb,twocolumn]{revtex4}       

\usepackage[dvips]{graphicx}
\input{epsf}

\begin{document}

\title{Resistive $g$-modes in a reversed field pinch plasma }

\author{M. Zuin$^{1}$, S. Spagnolo$^{1,2}$, R. Paccagnella$^{1}$, E. Martines$^{1}$, R. Cavazzana$^{1}$, G. Serianni$^{1}$, M. Spolaore$^{1}$, N. Vianello$^{1}$}
\affiliation{$^{1}$Consorzio RFX, Associazione EURATOM-ENEA sulla Fusione,
 35127 Padova, Italy\\$^{2}$Dipartimento di Fisica 'G. Galilei', Universit\`{a} degli Studi di Padova,  35131 Padova, Italy}


\begin{abstract}
First direct experimental evidence of high frequency, high toroidal mode number ($n>20$), magnetic fluctuations due to unstable resistive interchange modes ($g$-modes) resonant in the edge region of a reversed field pinch (RFP) plasma is presented.
Experimental characterization of time and space periodicities of the modes is provided by means of highly resolved in-vessel edge and insertable magnetic diagnostics.
It is found that the spectral mode properties are in good agreement with the predictions of the theoretical linear resistive magnetohydrodynamic stability analysis. A simple model is proposed for the observed saturation levels of the modes.
  Pacs: 52.25.Xz,52.30.Cv,52.35.Py,52.55.Hc,52.65.Kj,52.70.Ds

\end{abstract}

\maketitle

MHD turbulence of interchange nature is ubiquitous in both
astrophysical and laboratory plasmas. In fusion oriented magnetically confined plasmas interchange modes (also known as $g$-modes) can cause significant anomalous and particle transport across the magnetic field. In tokamak plasmas
$g$-modes are counteracted by
shaping and curvature effects of the mean field. The remaining
turbulence is therefore of ballooning nature \cite{Connor_PRL78} being
especially concentrated in the unfavorable toroidal curvature
region. For example, Edge Localized Modes (ELMs) are interpreted
in this way \cite{Tokar_PRL09}. In the reversed field pinch (RFP)
 \cite{Ortolani_Schnack} the average edge curvature is
unfavorable, being the field lines mainly in the poloidal
direction, therefore interchange modes should be expected.
According to Suydam's criterion \cite{Suydam} the effect of
unfavorable curvature can be mitigated by enhancing the magnetic
shear, and this is particularly relevant for RFPs. However, finite
plasma resistivity can give rise to a resistive interchange branch
of instabilities, known as resistive $g$-modes, that can be
destabilized well below the Suydam's limit. In the existing literature $g$-modes
have been often envisaged to be linearly unstable and invoked as
responsible for magnetic field ergodization and enhanced diffusion
and transport in RFP plasmas
\cite{Hender83,Carreras98,Manheimer_PRL80,Wyman_NF2009}. Some experimental results on this subject are described in ref. \cite{Brunsell_PoP94}. In this Letter, we
present the first clear experimental evidence of the existence
within the magnetic spectrum of resistive $g$-modes characterised
by high mode numbers along with a first direct comparison with modeling prediction.

The experimental activity here described has been performed on
RFX-mod \cite{Lorenzini_NP2009}, the largest operating toroidal
RFP device (major radius $R=2$ m, minor radius $a=0.459$ m).
Hydrogen is the working gas and the plasma current $I_p$ spans
between 0.3 and 1.8 MA. The on-axis electron temperature at the
highest plasma current exceeds 1.5  keV, while the volume averaged
electron density $n_e$ is in the range $1-10 \times  10^{19}
m^{-3}$. Discharge duration is up to 0.5 s. The analysis is
performed on a large number of discharges ($\sim$1000)  with
reversal parameter, defined as $F=B_{\phi}(a)/<B_{\phi}>$
($B_{\phi}(a)$ and $<B_{\phi}>$ toroidal field at the wall and
toroidal field averaged over the plasma cross-section,
respectively), in the range $[-0.5,0]$, which corresponds to edge
safety factor $q(a)\in[-0.07,0]$. The plasma is ohmically heated
and the loop voltage is
in the range 20 - 40 V,  which gives to the RFP the attractive feature of
being able to use ohmic heating only also in fusion-relevant
conditions. In the RFX-mod device the magnetic boundary is
determined by a thin Cu shell, with vertical field penetration
time of 50 ms, located at $r/a\cong1.12$, and by a system of 192
active saddle coils, supervised by a digital feedback system,
fully covering the machine with the aim of controlling the radial
fields due to field errors and MHD modes \cite{Sonato}. In the
experiments described herein the mesh of saddle coils has been
used in the so-dubbed  clean-mode-control (CMC) configuration, in
which the flux through each sensor loop is controlled by the
corresponding active coil after performing the two-dimensional FFT
of $B_r$ and $B_{\phi}$ measurements and computing a real time
correction of the aliasing of the sideband harmonics generated by
the discrete saddle coils \cite{Zanca_NF07}.

Magnetic
fluctuations have been measured by two different highly space and time resolved  systems of in-vessel magnetic probes: the first is an insertable edge probe, used to investigate high order toroidal harmonics ($|n|\leq85$), the second consists of arrays of probes, covering the full toroidal and poloidal angles of the torus.
The insertable probe, named \textit{U-probe} \cite{Zuin_PPCF2009}, which was used in a set of experiments performed at low toroidal plasma current
($I_{p} \leq 500$ kA), consists of two Boron Nitride cases, 5 cm toroidally spaced, radially inserted from the vacuum
chamber up to $r/a \approx  0.9$ at $\theta=0^{\circ}$ (defined on the equatorial plane on the low field side of the machine with positive $\theta$ pointing in the upward direction), without significant perturbation
of the plasma. Each case contains, along with a number of electrostatic pins,  a radial array of 7 three-axial magnetic
pick-up coils measuring the time
derivative of the three components of the magnetic field ($\dot{B}_{r}$, $\dot{B}_{\theta}$, $\dot{B}_{\phi}$). The sampling frequency is 10 MHz, with an estimated bandwidth up to 3 MHz.
\begin{figure}
\centering
\includegraphics[width=0.86\columnwidth]{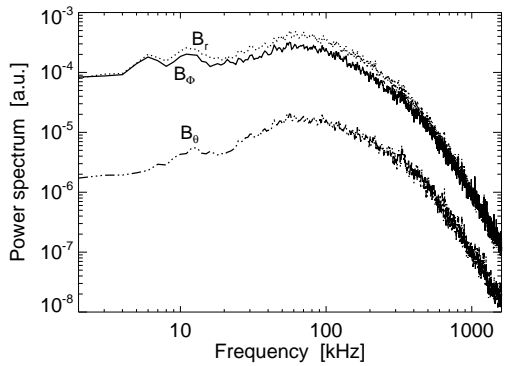}
\caption{Typical power spectra of the time derivative of the three components of the magnetic field, measured by one of the three-axial coils housed in the \textit{U-probe}.} \label{Sf}
\end{figure}
The second system of probes is a subset of the ISIS (Integrated System of Internal Sensors) diagnostics \cite{Serianni_ISIS}, a large set of high frequency electrostatic and magnetic probes located inside the vacuum vessel. The magnetic probes consist of coils measuring $\dot{B}_{\phi}$, placed at $r/a=1.03$ behind the graphite tiles which cover the first wall of the machine. The probes used in this study are distributed in two different arrays: the first constituted of 48 coils evenly distributed in the toroidal direction located at $\theta=250^{\circ}$; the second one is a poloidal array, made of 8 equally spaced coils.
The sampling frequency for the magnetic measurements from the ISIS system is 2 MHz, while the estimated bandwidth is up to 400 kHz.
The two arrays allow a resolution of the toroidal and poloidal mode numbers $n$ and $m$ up to 24 and 4, respectively.

\begin{figure}
\centering
\includegraphics[width=0.86\columnwidth]{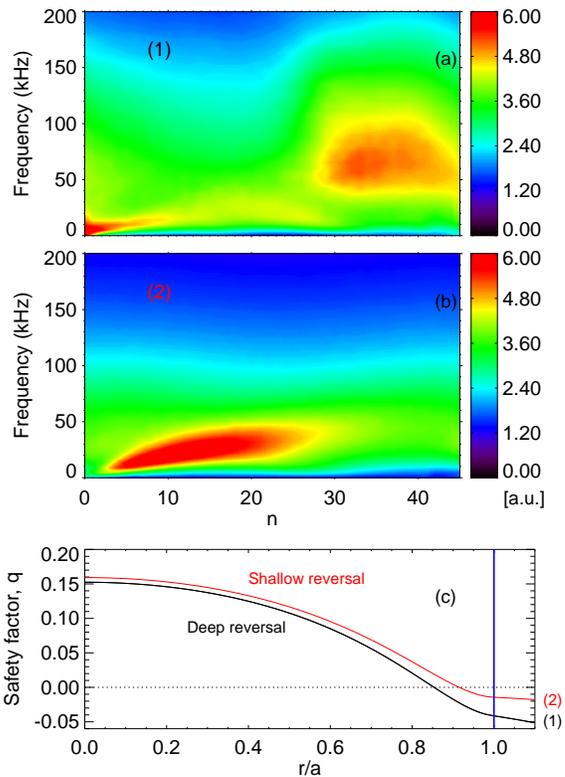}
\caption{Typical color coded contour plot of the $S(n,f)$ spectra of magnetic fluctuations for two different experimental conditions:  a)  $S(n,f)$ spectrum obtained at $q(a)\approx -0.05$;  b)  $S(n,f)$ spectrum obtained at shallower reversal of the magnetic field ($q(a)\approx -0.015$); c) the two reference $q(r)$ profiles (the vertical blue line marks the plasma edge). In the sets of discharges considered $300kA<I_p<600kA$.} \label{Snf_2qa}
\end{figure}

In Fig. \ref{Sf} the typical power spectra of the signals from the three components of the magnetic field fluctuations ($\dot{B}_r$, $\dot{B}_{\theta}$, $\dot{B}_{\phi}$), taken with the \textit{U-probe} in a deeply reversed ($q(a)\approx -0.04$) discharge, are shown. The spectra show that the highest fluctuation levels are present in the ($\dot{B}_r$,$\dot{B}_{\phi}$) signals, which correspond to the perpendicular components at the edge of the RFP configuration, where the dominant magnetic field is ${B}_{\theta}$. The spectra show that large magnetic fluctuations are present for frequencies up to about 200 kHz, with a large part of the magnetic fluctuating energy concentrated in the frequency region spanning from 30 to 150 kHz.  
The space-time properties of the magnetic fluctuations have been  obtained by Fourier decomposing the signals coming from the complete toroidal array of coils measuring $\dot{B}_{\phi}$. 
In Fig. \ref{Snf_2qa} the frequency spectrum of each toroidal Fourier component is plotted in a color-coded plot for two different experimental operating conditions, corresponding to plasma equilibria characterised by the $q$ profiles plotted in Fig. \ref{Snf_2qa}c. The $q$ profiles are deduced by using a suitable magnetic profile reconstruction (see for example \cite{Paccagnella_NF98}). Each of the spectra has been obtained as an average over about 200 discharges.  Fig. \ref{Snf_2qa}a has been obtained by analysing data taken in a set of discharges with deep reversal of the toroidal magnetic field ($q(a)\approx-0.05$); the $S(n,f)$ spectrum exhibits the excitation of coherent magnetic fluctuations, in the form of a broad peak, at frequencies above 30 kHz and high ($n$ up to 50) positive toroidal mode numbers (in the RFX-mod convention, negative and positive $n$ values refer to modes resonant inside and outside the reversal surface, respectively).
It is worth noting that due to the limited ($|n|\leq 24$) spatial resolution of the toroidal array of coils, the computed $S(n,f)$ spectra, originally affected by aliasing, had to be corrected. This operation was made possible by a comparison to the $S(n,f)$ spectra deduced by applying the two-point technique \cite{Power} to the signals from the closely spaced magnetic coils of the \textit{U-probe}, which show that the observed peak actually correspond to high positive $n$ values.
Fig. \ref{Snf_2qa}b shows the result of the same analysis performed on discharges characterised by a shallow reversal of the toroidal magnetic field, $q(a)\approx -0.015$. In this case no peak is present and the magnetic fluctuations are concentrated at frequencies below 50 kHz, positioned on an almost continuous linear dispersion relation, in the form $\omega=\frac{n}{R} V_{ph}$, with $\omega=2\pi f$, where $f$ is the frequency, $n$ is the toroidal mode number, and $V_{ph}$ is the phase velocity of the fluctuations in the toroidal direction. The measured phase velocity is of the order of 20-30 km/s in the counter toroidal plasma current direction, which is in good agreement with the plasma flow velocity at the edge of the RFX-mod device deduced by means of a gas puffing imaging diagnostics \cite{Scarin_JNM07} and is also consistent with the $E \times B$ flow measured by Langmuir probes \cite{Spolaore_PRL2009}.
\begin{figure}
\centering
\includegraphics[width=.86\columnwidth]{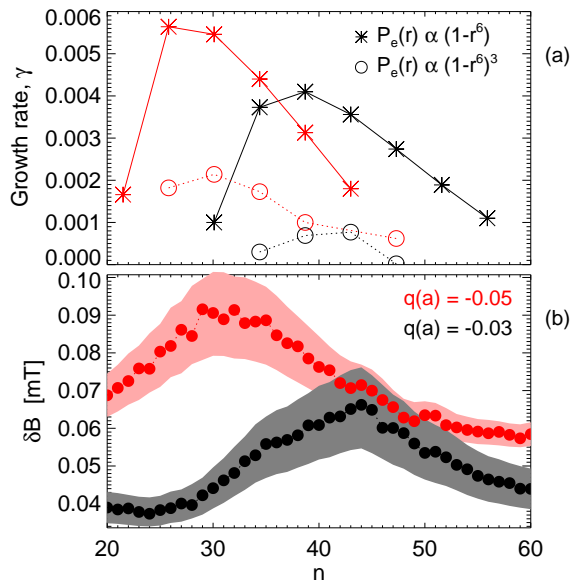}
\caption{a) Theoretical growth rate for different toroidal mode numbers, corresponding to two $q(a)$. Different symbols mark the two different plasma pressure profiles used for the simulation. b) Experimental $S(n)$ spectra in the same $q(a)$ conditions of a) (experimental spectra are obtained as average over 10 comparable discharges, in terms of $I_p$, $n_e$ and $q(a)$). } \label{Sn_exp_and_theory}
\end{figure}
It is important to note that the peak in the $(n,f)$ plane of Fig. \ref{Snf_2qa}a results to be almost aligned with the linear dispersion relation characterising the continuum low frequency part of the spectrum. This is an indication that the frequency associated to the mode with $n>20$,  is mainly due to a Doppler effect, as the mode rotates with the plasma. 
The analysis for different values of $q(a)$ reveals that the value of
$n$ corresponding to the maximum of the fluctuation amplitude is observed to follow the relation $n\cdot|q(a)|\approx$1.
The analysis of the poloidal periodicities of the magnetic signals from the poloidal array of coils reveals a clear $m=1$ nature of the magnetic fluctuations in the explored frequency range.
The measured poloidal and toroidal mode numbers thus indicate that the observed high frequency modes correspond to
magnetic perturbations resonating at the edge of the plasma column, in the region between the reversal surface and the inner surface of the graphite tiles constituting the first wall of the RFX-mod machine. In the RFP nomenclature these are called externally resonant modes \cite{Paccagnella_NF98}.

The measured spectra have been compared to those predicted by a
linear stability analysis performed by means of the cylindrical
code ETAW, already extensively used for RFP calculations
\cite{Paccagnella_NF98} and more recently successfully benchmarked
against the MARS code \cite{villone08}. The code solves the linear
cylindrical resistive incompressible and inviscid single fluid MHD
equations, using a spectral formulation and a matrix shooting
eigenvalue scheme. Up to two resistive walls are considered for
the boundary conditions, with a thin shell approximation. The
plasma model is solved inside the first wall and the solution is
then matched to the external solution of the vacuum cylindrical
Laplace equation, analytically known in terms of Modified Bessel
Functions. For the problem under consideration we have assumed
that only one wall is present, which corresponds to a perfect ideal
shell at a shell proximity ($b/a$) of about 1.05, i.e. 5 $\% $  of
the plasma minor radius. This choice appears to be the most
appropriate, since, as it has been shown by the measurements, the
modes rotate at relatively high speed together with the plasma,
and, on the other hand, a metal liner with 1-2 ms magnetic field
penetration time, is present in the RFX-mod device. Therefore the
liner acts as a perfect conductor for these fast rotating modes.
\begin{figure}
\centering
\includegraphics[width=.86\columnwidth]{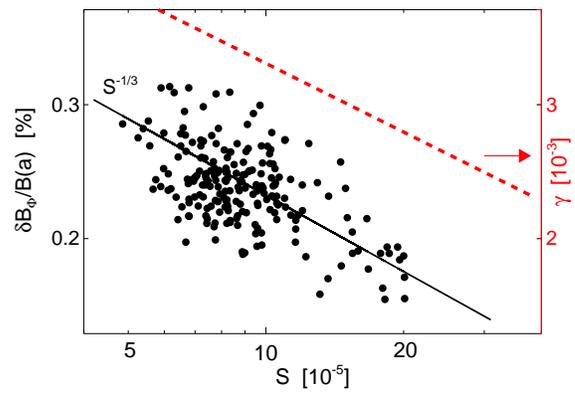}
\caption{Total normalised amplitude of the magnetic perturbation as a function of the Lundquist number $S$ for a given plasma equilibrium ($q(a)=-0.03$). Dashed line, referring to the right hand side y-axis, represents the dependence on $S$ of the growth rate predicted by the ETAW code for the same plasma equilibrium. } \label{rms_Bt_vs_S}
\end{figure}

With these assumptions, we obtained the spectrum of unstable modes given in Fig. \ref{Sn_exp_and_theory}a for two different $q$ profiles, one with a deeper reversal, corresponding to $q(a)=-0.05$, in which the
spectrum is peaked at a lower $n$ value ($n \approx 25$), and one at
a shallower reversal, with $q(a)=-0.03$ , with a spectrum peak at $
n \approx 40 $.
The $\beta$  used in the simulation is 2\%, comparable to the
experimental one for the considered case. The theoretical
predictions are in good agreement with the experimental toroidal
mode number spectra $S(n)$, shown in Fig.
\ref{Sn_exp_and_theory}b, obtained by integrating the $S(n,f)$
spectra over frequencies above 30 kHz, which, as discussed above,
are those pertaining to the investigated $m=1$ modes (in order to
obtain the value of the fluctuating magnetic field, magnetic
signals have been numerically integrated). In particular,
similarly to the experimental results, the toroidal mode number
spectrum for a given equilibrium of Fig. \ref{Sn_exp_and_theory}a
shows that the growth rate $\gamma$ (normalised to the
Alfv\'{e}n time) vanishes for modes resonant close to the
stabilising wall (at low $n$'s), and also for modes resonant close
to the reversal of the toroidal magnetic field (at high $n$'s),
where the effect of the magnetic shear is stronger.
As also shown in Fig. \ref{Sn_exp_and_theory}a, the theoretical
growth rates at a given $\beta$ are strongly influenced by the
pressure profile. In particular, larger edge pressure gradients
are associated to larger $\gamma$ values. Moreover, modes are
found to be linearly stable at zero $\beta$.
The model also indicates that the growth rates scale like $\sim
S^{-0.3}$ ($S$ being the Lundquist number), as shown in Fig.
\ref{rms_Bt_vs_S}, for a given plasma equilibrium corresponding to
$q(a)=-0.03$, which is very close to the theoretically expected
$S^{-1/3}$. In the same figure, the theoretical growth rates are
compared to the experimental total amplitude of the magnetic
fluctuations, obtained by summing the contribution from each
single toroidal harmonic. Despite the large dispersion of the
experimental points, a scaling close to $S^{-1/3}$ is fully
compatible with the data.

For all these reasons, i.e. effect of the resonance, effect of the
beta and shear and Lundquist scaling, we can classify these modes
as resistive $g$. These are interchange-like instabilities which
can develop in a RFP at relatively low $\beta $ values
\cite{merlin}, well below the Suydam's limit for ideal
interchanges.

\begin{figure}
\centering
\includegraphics[width=.86\columnwidth]{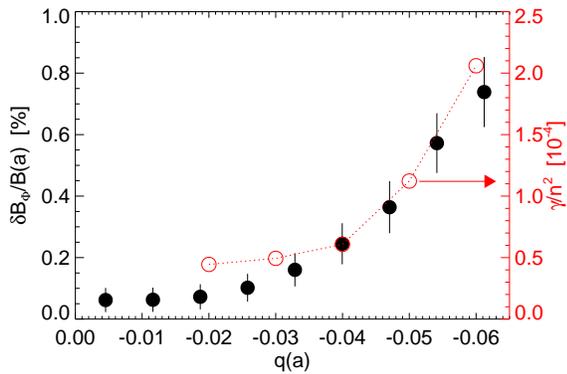}
\caption{Total (normalised) magnetic perturbation $\delta B_{\phi}$ induced by the resistive $g$-modes as a function of $q(a)$ (full circles). Theoretical growth rate, normalized to the squared $n$ values as a function of $q(a)$ (open circles).} \label{rms_Bt_vs_qa}
\end{figure}

As a last point we want to address the problem of
the nonlinear saturation of these modes. Clearly for this purpose
a nonlinear model would be required (see for example \cite{battha}), and/or a model taking into account all the effects which could affect the stability properties of $g$-modes extensively described in the literature, with the inclusion of finite Larmor effects, parallel ion viscosity, diamagnetics effects \cite{Sydora,Zhu} (and references therein).
However, we consider a simple model in which the growth rate is
balanced by a dissipation mechanism proportional to the squared
mode wave-number, which is the consequence of having a
Laplacian-like dissipation term in the equations. In this case, it
is expected that the measured fluctuation amplitude correlates with some quantity which takes into account both the
growth rate and the mode number. As shown in Fig.
\ref{rms_Bt_vs_qa}, despite the semplicity of the model, a
very nice correlation is found between the maximum growth rate
divided by the square of the mode number (corresponding to the
maximum) at different $q(a)$ values and the experimental amplitude
of the fluctuations. 
In particular, in Fig. \ref{rms_Bt_vs_qa} the
amplitude of the total perturbation produced by these modes is
observed to reach values up to almost 1\% of the equilibrium
magnetic field at the edge in deeply reversed discharges. It is
worth to note that the magnetic perturbation produced is only a
factor four smaller than that due to the core resonant dynamo
modes, which are essential to sustain the RFP configuration
\cite{Lorenzini_NP2009}. Therefore we think that the described phenomenon could play an important role in determining the edge transport properties in RFP plasmas, for which the exact mechanism is still under debate.

\emph{This work, supported by the European Communities under the contract
of Association between EURATOM/ENEA, was carried out within the
framework the European Fusion Development Agreement.} 



%

\end{document}